\documentclass[ amsmath,amssymb,aps,prb,reprint,superscriptaddress,showpacs]{revtex4-1}

\usepackage{graphicx}
\usepackage{dcolumn}
\usepackage{bm}
\usepackage{amsmath}

\begin{document}

\title{Multiband effect in elastoresistance of Fe(Se,Te)}

\author{Y.A.~Ovchenkov}
\affiliation{Faculty of Physics, M.V. Lomonosov Moscow State University, Moscow 119991, Russia}
\author{D.A.~Chareev}
\affiliation{Institute of Experimental Mineralogy, RAS, Chernogolovka, 123456, Russia}
\affiliation{Ural Federal University, Ekaterinburg, 620002, Russia }
\affiliation{Institute of Geology and Petroleum Technologies, Kazan Federal University, Kazan, 420008, Russia }
\author{D.E.~Presnov}
\affiliation{Faculty of Physics, M.V. Lomonosov Moscow State University, Moscow 119991, Russia}
\affiliation{MSU Quantum Technology Centre, Moscow, 119991, Russia}
\affiliation{Skobeltsyn Institute of Nuclear Physics, Moscow 119991, Russia}

\author{I.G.~Puzanova}
\affiliation{State University "Dubna", Dubna, 141980, Russia}

\author{O.S.~Volkova}
\affiliation{Faculty of Physics, M.V. Lomonosov Moscow State University, Moscow 119991, Russia}
\affiliation{Ural Federal University, Ekaterinburg, 620002, Russia }
\affiliation{National University of Science and Technology `MISiS,' Moscow 119049, Russia }

\author{A.N.~Vasiliev}
\affiliation{Faculty of Physics, M.V. Lomonosov Moscow State University, Moscow 119991, Russia}
\affiliation{Ural Federal University, Ekaterinburg, 620002, Russia }

\date{\today}
%
\begin{abstract}

We have investigated the elastoresistance of two FeSe${}_{1-x}$Te${}_{x}$ (x about 0.4 - 0.5) compounds that have a close chemical composition but differ significantly in electronic properties.
The first compound has a negative temperature coefficient of resistance and does not show any phase transitions other than superconducting. The elastoresistance of this compound approximately follows $1/T$ low as it usually occurs in Fe(Se,S) with metallic conductivity. The second compound has a metallic type of conductivity and in addition to the superconducting transition, there is also a phase transition at a temperature of about 30 K. The elastoresistance of the second compound is sign-reversing and can be approximated with the sum of two Curie-Weiss type terms with opposite signs and different critical temperatures which suggest a competition of contributions to the elastoresistance from different band valleys. 

\end{abstract}
%
%
\maketitle
%
%
%
\section{Introduction}

The high values of the strain coefficient of electrical resistivity are observed in many series of iron-based superconductors. This interesting phenomenon is being actively studied now since the reasons for the large elastoresistance effect can be associated with the mechanism of superconducting pairing in this family of superconductors \cite{chu2012, Bohmer2018,BOHMER201690}.

In the semiclassical approximation, the contribution to a conductivity from the $i$-th group of carriers $\sigma_{i}$ is determined by the concentration of carriers $n_{i}$, the effective mass of carriers $m^{*}_{i}$, and the scattering time $\tau_{i}$:
\begin{eqnarray}
\sigma_{i}=\frac{en_{i}\tau_{i}}{m^{*}_{i}}
\label{R_eq}
\end{eqnarray}

The elastoresistance of many ordinary metals and alloys is reasonably well explained \cite{PhysRev.94.61} under the assumption that the effective mass does not change during deformation and the density of states and scattering time change only due to a change in volume. In this approach, changes in concentration and scattering time provide roughly equal contributions to the strain coefficient of resistivity which are of the order of unity.

For silicon, the gauge factor can reach values of the order of a hundred. In semiconductor materials, the change in the value of resistivity during deformation is primarily due to the redistribution of carriers between the valleys with different anisotropy \cite{PhysRev.94.42}. In other words, due to changes in the concentrations $n_{i}$ of the valleys. Change in the anisotropy of one of the valleys is also an important factor for silicon \cite{kanda1991}.

In semimetals, the elastoresistance can also be large. For bismuth, for example, this value is of the order of a ten. Recently, values of the order of a hundred have been reported for the elastoresistance of one of the transition metal dichalcogenides \cite{Jo25524}.

In the case of iron-based superconductors, the magnitude of the elastoresistance can reach several hundred. The temperature dependence of the elastoresistance is often of the Curie-Weiss type with diverging behavior at the points of structural or magnetic transitions, which suggest the electronic origin of the corresponding transitions. An important question so far remains which of the properties of the electronic subsystem causes this divergence. In the quasiclassical variables of equation (\ref{R_eq}) this can be either the thermodynamic properties of the carriers — $m^{*}_{i}$ and $n_{i}$ or the scattering time $\tau_{i}$. The last of these possibilities is of great interest since it can imply a significant role of nematic fluctuations, which may also be important for understanding the mechanism of superconducting pairing in iron-based superconductors.

On the other hand, large values of elastoresistance can also be explained in the framework of the band approach, taking into account only the thermodynamic properties of carriers. This approach is confirmed by recent experimental results  \cite{ovchenkov2019nematic}. For example, it was found that a change in the sign of elastoresistance in series 1-1 occurs in the same composition range where a change in the sign of the main carriers is expected \cite{Ovchenkov_2019}. A change in the sign of resistance anisotropy in the ordered state in compositions 1-2-2 also correlates with the type of doping \cite{blomberg2013sign}.

The noticeable effect of strain on the electronic structure in FeSe was directly observed experimentally \cite{PhysRevB.95.224507}. It is also well known that the band structure in FeSe${}_{1-x}$Te${}_{x}$ changes significantly during the nematic transition \cite{Coldea2019} and the effective mass of carriers in the nematic phase is anisotropic for some valleys. Thus, the band contribution to elastoresistance certainly exists.

In general, the components of the mass tensor and the relaxation time tensor enter the Boltzmann equation as pairwise products of the relaxation time and the reciprocal mass. Therefore, within the framework of the semiclassical approximation, the separation of the contribution of these quantities to transport properties is ambiguous. Nevertheless, the question of the sources of the record values of elastoresistance is as important as the question "what drives nematic order in iron-based superconductors?" \cite{Fernandes2014}. 

To study the relationship between the elastoresistance and other electronic properties, we investigated the properties of two FeSe${}_{1-x}$Te${}_{x}$ compounds with close $x$ but with the opposite signs of the temperature coefficient of resistivity. In our opinion, the results obtained indicate a close relationship between the elastoresistance and band properties. In particular, a competition of contributions from different valleys leads to the temperature dependence of the elastoresistance with a change in sign found in one of the compounds studied.

\section{Experiment}
There were selected two compositions from a rather large number of studied batches of the FeSe${}_{1-x}$Te${}_{x}$ series for x near 0.4-0.5, as having a maximum difference in transport properties, presumably corresponding to the boundary values of the deviation of the stoichiometry of iron. The range of values of $x$ is chosen because of a transition from bad to good metal occurs nearby, and samples with almost the same value of $x$ can have different signs of the temperature coefficient of resistance. It is also worth noting that in this range the conditions for topological superconductivity are fulfilled \cite{Zhang182}.

Single crystals of Fe${}_{1+\Delta}$Se${}_{0.5}$Te${}_{0.5}$ and  Fe${}_{1+\delta}$Se${}_{0.6}$Te${}_{0.4}$ were grown using recrystallization in halides flux technique with a constant temperature gradient along quartz ampoule. A driving force of the recrystallization process is the temperature gradient. The mixture of Fe, Se, and Te powders was dissolved in molten salts at the hot end of an ampoule and transferred to the cold one where crystallization occurs. The difference between temperatures of hot and cold ends of the ampoule was about 50-100K. See ref. \cite{chareev2016general, chareev2016synthesis} for more details of this technique.

Sample Fe${}_{1+\Delta}$Se${}_{0.5}$Te${}_{0.5}$ was synthesized in CsCl/KCl/NaCl eutectic flux, the temperature of the hot end of the ampoule was 655$^\circ$C and of the cold end was 575$^\circ$C. Sample Fe${}_{1+\delta}$Se${}_{0.6}$Te${}_{0.4}$ was synthesized in AlCl$_{3}$/NaCl/KCl flux and the temperature mode was 585$^\circ$C at the hot end and 495$^\circ$C at the cold end. Both syntheses were carried out for about 75 days. 

The chemical composition of crystals was studied with a digital scanning electronic microscope TESCAN Vega II XMU with the energy dispersive microanalysis system INCA Energy 450 (accelerating voltage 20 kV, probe current 0.4 nA). The experimentally determined chemical composition values were Fe${}_{1.02}$Se${}_{0.49}$Te${}_{0.51}$ and Fe${}_{0.98}$Se${}_{0.6}$Te${}_{0.4}$ for the batches designated as Fe${}_{1+\Delta}$Se${}_{0.5}$Te${}_{0.5}$ and  Fe${}_{1+\delta}$Se${}_{0.6}$Te${}_{0.4}$, respectively.

\begin{figure}[h]
\centering
  \includegraphics[scale=0.5]{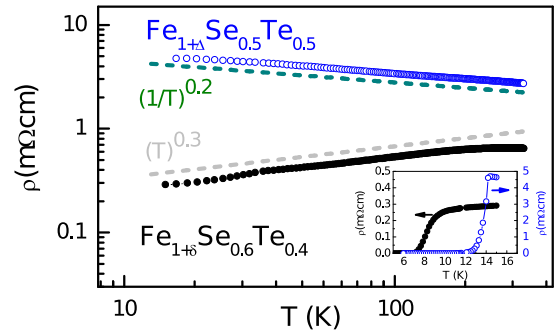}
  \caption{Log-log plot of the temperature dependence of resistivity for Fe${}_{1+\delta}$Se${}_{0.6}$Te${}_{0.4}$ and Fe${}_{1+\Delta}$Se${}_{0.5}$Te${}_{0.5}$ above 15 K. For reference, two dashed lines with a slope of -0.2 and 0.3 are shown. Inset: Temperature dependence of resistivity below 15 K. }
  \label{fgr:fig1}
\end{figure}

Contact pads for electrical measurements were made by magnetron sputtering of Au/Ti layers using a mechanical mask. 
DC transport measurements were done using QD PPMS and EDX options of MPMS 7T with Keithley 2400 and Keithley 2192. Elastoresistivity was measured similarly to the method described in ref. \cite{chu2012} using AC transport option of a Quantum Design PPMS system equipped with a multifunctional insert. During measurements, the sample was glued to a commercial piezoelectric transducer. The sample elongation was measured by a strain gauge located on the other side of the piezoelectric device.
%
\section{Results}

The temperature dependences of the resistivity of the studied compositions are shown in  Fig.~\ref{fgr:fig1}. In the log-log plot, the experimental data are almost linear in a wide temperature range. For Fe${}_{1+\Delta}$Se${}_{0.5}$Te${}_{0.5}$ the slope is about -0.2, which is close to the Mott law and apparently reflects the hopping nature of the conductivity in the studied composition. This is in agreement with the supposed increased excess of iron. The critical temperature for this composition is close to 14 K. One should notice that a negative value of the temperature coefficient is almost standard for compositions with 50\% tellurium and was observed by many authors \cite{JPSJ.79.102001}, especially in the early years of studying this family. 

It is known that slight variations in the synthesis of compounds of the family 1-1 can lead to rather noticeable changes in the transport properties of crystals \cite{bohmer2016variation}. It was also found that the transport properties noticeably change with various heat treatments that remove non-stoichiometric iron \cite{Sun_2019}.

For our Fe${}_{1+\delta}$Se${}_{0.6}$Te${}_{0.4}$ the slope of R(T) in log-log plots is about 0.3. This is closer to the metallic behavior observed in FeSe${}_{1-x}$S${}_{x}$ \cite{2017_Coldea,Char_FeSeS_CEC}, although an activation component is also present in the resistance of the FeSe at high temperatures \cite{karlsson2015study}.

In general, the temperature dependence of the resistance of Fe${}_{1+\delta}$Se${}_{0.6}$Te${}_{0.4}$ agrees very well with the recently published data for Fe${}_{1+\delta}$Se${}_{1-x}$Te${}_{x}$ compositions \cite{PhysRevB.100.224516}, including an anomaly in resistance at a temperature of about 30 K, presumably caused by a magnetic transition.

\begin{figure}[h]
\centering
  \includegraphics[scale=0.5]{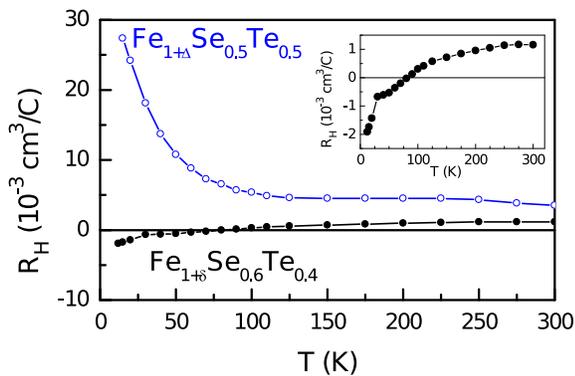}
  \caption{Temperature dependence of the Hall coefficient for Fe${}_{1+\delta}$Se${}_{0.6}$Te${}_{0.4}$ and Fe${}_{1+\Delta}$Se${}_{0.5}$Te${}_{0.5}$. Inset: Temperature dependence of the Hall coefficient for Fe${}_{1+\delta}$Se${}_{0.6}$Te${}_{0.4}$. }
  \label{fgr:fig2}
\end{figure}

The temperature dependences of the Hall coefficients of the compositions studied are shown in Fig.~\ref{fgr:fig2}. This plot also indicates a significant difference in the electronic properties of the samples. For Fe${}_{1+\delta}$Se${}_{0.6}$Te${}_{0.4}$ the dependence again similar to the case of metallic FeSe${}_{1-x}$S${}_{x}$ where the Hall coefficient has a change of sign caused by the competition of electronic and hole components. For Fe${}_{1+\Delta}$Se${}_{0.5}$Te${}_{0.5}$ $R_{H}$ is positive at all temperatures. It is known that heat treatments of  FeSe${}_{1-x}$Te${}_{x}$, which remove part of the excess of iron, also cause a change in the dependencies of the Hall coefficient from always positive to dependence with a change of sign \cite{sun2015evolution}. Thus, the substantial difference in the transport properties of the studied compounds with similar composition can be explained by variations in the non-stoichiometry of iron. The excess non-stoichiometric iron in Fe${}_{1+\Delta}$Se${}_{0.5}$Te${}_{0.5}$ dopes this composition and increases the degree of disorder.

The difference in the electronic transport properties of our compounds also manifests itself in elastoresistance (see Fig.~\ref{fgr:fig3}). Elastoresistance of Fe${}_{1+\Delta}$Se${}_{0.5}$Te${}_{0.5}$ slowly increases with the decreasing temperature down to about 50 K, where it shows saturation. This dependence is fairly well approximated by the $A/T$ term and constant in almost the entire temperature range. The absolute values of elastoresistance of this composition do not exceed 20.

In the sample Fe${}_{1+\delta}$Se${}_{0.6}$Te${}_{0.4}$, the behavior of elastoresistance is much more complicated. There are two temperature changes in the sign of the elastoresistance. Nevertheless, above the temperature of the structural transition, the  dependence of the elastoresistance is well approximated by two terms of the  $A/(T-\theta)$ type with different signs.


\section{Discussion and conclusion}

\begin{figure}[h]
\centering
  \includegraphics[scale=0.5]{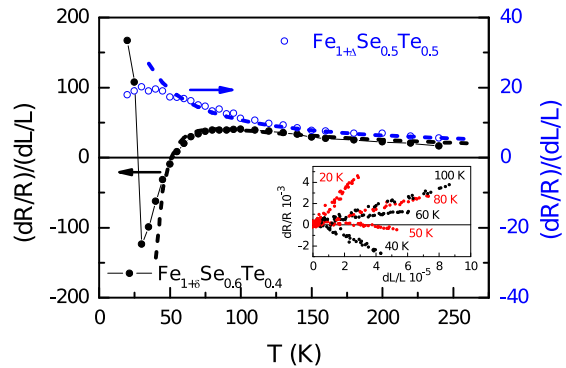}
  \caption{Temperature dependence of the longitudinal elastoresistivity $(\Delta{}\rho_{xx}/\rho_{xx})/(\Delta{}l_{x}/l_{x})$ for Fe${}_{1+\delta}$Se${}_{0.6}$Te${}_{0.4}$ and Fe${}_{1+\Delta}$Se${}_{0.5}$Te${}_{0.5}$. The dotted curves represent a Curie-Weiss type dependence and the sum of two Curie-Weiss terms with different parameters. ($870/T+2$ and $9000/T-3700/(T-30)+2$) Inset: The relative change of resistivity $(\Delta{}\rho_{xx}/\rho_{xx})$ as a function of elongation $(\Delta{}l_{x}/l_{x})$ at several temperatures. }
  \label{fgr:fig3}
\end{figure}

Depending on what is the source of elastoresistance, we expect different changes in the magnitude of the effect due to changes in electronic properties, for example, in scattering time. For example, if the cause of the elastoresistance is a band parameter $m^{*}$, then from equation (\ref{R_eq}) we can obtain:

\begin{eqnarray}
\frac{1}{\rho}\frac{\partial{\rho}}{\partial{\epsilon}}=\frac{1}{m^{*}}\frac{\partial{m^{*}}}{\partial{\epsilon}}
\end{eqnarray}

where $\epsilon$ is elongation. Thus, in the case of the band mechanism of the elastoresistance, a change in the scattering time, for example, with an increase in disorder, does not change the behavior of elastoresistance.

If the elastoresistance is caused by a change in the scattering, then the elastoresistance will change in a completely different way when the scattering is modified. The elastoresistance should be completely suppressed when scattering by nematic fluctuations becomes insignificant in comparison with other contributions to scattering. For samples "a" and "b" differing only in scattering, in the approximation of independent scattering times, from equation  (\ref{R_eq}) we can obtain:

\begin{eqnarray}
\frac{1}{\rho_{a}}\frac{\partial{\rho_{a}}}{\partial{\epsilon}}=\frac{1}{\rho_{b}}\frac{\partial{\rho_{b}}}{\partial{\epsilon}}\times(\frac{\rho_{b}}{\rho_{a}})
\label{dR_eq}
\end{eqnarray}

For the two compounds studied, the ratio of resistance is close to the temperature function in $0.5$ power. Moreover, for FeSe${}_{1-x}$S${}_{x}$  the resistance increases with temperature faster than the linear function in a wide temperature range \cite{SUST-30-3-035017}. But in all these cases, the experiments reveal $ T^{-1}$ dependencies for elastoresistance. This indicates that the relation (\ref{dR_eq}) is most likely not applicable.

We also consider the complicated temperature dependence of the elastoresistance for Fe${}_{1+\delta}$Se${}_{0.6}$Te${}_{0.4}$ to be a manifestation of the band nature of the elastoresistance. We attribute this behavior to the competition of contributions from different valleys as will be explained below.

For two independent group of carriers the total resistance has the expression:

\begin{eqnarray}
\rho_{1+2}=\frac{\rho_{1}\rho_{2}}{(\rho_{1}+\rho_{2})}
\label{R_sum}
\end{eqnarray}

where $\rho_{1+2}$ is total resistance, and $\rho_{1,2}=1/\sigma_{1,2}$ is a reciprocal of the conductivity of one group. Elastoresistance will be expressed as:

\begin{eqnarray}
\frac{1}{\rho_{1+2}}\frac{\partial{\rho_{1+2}}}{\partial{\epsilon}}=\frac{1}{\rho_{1}}\frac{\partial{\rho_{1}}}{\partial{\epsilon}}\frac{\rho_{2}}{\rho_{1}+\rho_{2}}+\frac{1}{\rho_{2}}\frac{\partial{\rho_{2}}}{\partial{\epsilon}}\frac{\rho_{1}}{\rho_{1}+\rho_{2}}
\label{dR_sum}
\end{eqnarray}

If ratio $\sigma_{1}/\sigma_{2}$ is temperature independent, which is a good approximation in many cases, then the elastoresistance of the material can be represented as a linear combination of the contributions of two groups.

It should be recognized that the sum of two independent contributions can be obtained under various assumptions. For example, such a sum can describe the contributions of different phases in the presence of phase separation or bulk and surface contributions. Compositions near FeSe${}_{0.6}$Te${}_{0.4}$ with metallic conductivity and magnetic transition little has been studied. Perhaps they have features not yet known. But now the most probable is the assumption of the microscopic nature of two contributions in the elastoresistance.

As it can be seen from Fig.~\ref{fgr:fig3}, the temperature dependence of the elastoresistance of sample  Fe${}_{1+\delta}$Se${}_{0.6}$Te${}_{0.4}$ is satisfactorily described by the sum of two Curie-Weiss-type contributions with different signs and different critical temperatures. The difference in critical temperatures is a key feature since it provides a change of sign. In the band model of elastoresistance, different values of critical temperatures for different valleys are a completely clear assumption, since it can be a local characteristic of the valley. It seems more sophisticated to find the reason for the existence of two different critical points for scattering processes.

In conclusion.
The study of elastoresistance of two close compositions FeSe${}_{1-x}$Te${}_{x}$ with substantially different electronic properties revealed two important features. First, a change in the type of conductivity of the compositions does not change the type of temperature dependence of elastoresistance. Second, the sign-reversal temperature dependence of the elastoresistance is observed, which is explained by the existence of two competing contributions with different critical temperatures.

Both detected phenomena have a simple explanation in the framework of the band model of elastoresistance. Thus, the record values of the elastoresistance along with the record values of the absorption of ultrasound \cite{epl_acustic} and the record increase of the superconducting temperature under pressure \cite{Medvedev2009} can be attributed to the features of the band structure of iron-based superconductors.

\begin{acknowledgments}
This work was supported in part from the Ministry of Education and Science of the Russian Federation in the framework of Increase Competitiveness Program of NUST 'MISiS' (K2-2015-075 and K4-2015-020) and by Act 211 of the Government of Russian Federation, agreement 02.A03.21.0006 and by the Russian Government Program of Competitive Growth of Kazan Federal University. We acknowledge support from Russian Foundation for Basic Research (Grants 20-02-00561 and ofi-m 17-29-10007) and Russian Science Foundation (Grant 19-42-02010).
\end{acknowledgments}
\bibliographystyle{unsrtnat}
\bibliography{FeTeSe_str.bib}
\end{document}